\documentclass[aps,prc,,12pt,preprint,showpacs,preprintnumbers,superscriptaddress]{revtex4-1}
\usepackage{graphicx}
\usepackage{dcolumn}
\usepackage{bm}
\usepackage{amsmath} 

\begin{document}

\title{Exact Eigenvalues of the Pairing Hamiltonian Using Continuum Level Density}
\author{R. Id Betan}
\affiliation{Department of Physics and Chemistry (FCEIA-UNR) -
             Physics Institute of Rosario (CONICET),\\
             Av. Pellegrini 250, S2000BTP Rosario, Argentina}

\date{\today}

\begin{abstract}
The pairing Hamiltonian constitutes an important approximation in many-
body systems, it is exactly soluble and quantum integrable. On the 
other hand, the continuum single particle level density (CSPLD) contains 
information about the continuum energy spectrum. The 
question whether one can use the Hamiltonian with constant pairing strength 
for correlations in the continuum is still unanswered. In this paper we 
generalize the Richardson exact solution for the pairing Hamiltonian 
including correlations in the continuum. 
The resonant and non-resonant continuum are included through the CSPLD. The 
resonant correlations are made 
explicit by using the Cauchy theorem. Low lying states with seniority zero 
and two are calculated for the even Carbon isotopes. We conclude that 
energy levels can indeed be calculated with constant pairing in the 
continuum using the CSPLD. It is found that the nucleus $^{24}$C 
is unbound. The real and complex energy representation of the continuum is 
developed and their differences are shown. The trajectory of the pair energies 
in the continuum for the nucleus $^{28}$C is shown.
\end{abstract}


\maketitle

\section{Introduction} \label{sec.introduction}
The approximate BCS solution of the pairing Hamiltonian has been extensively used in Condensed Matter to study pairing correlations in ultra-small metallic grains \cite{1996VonDelft,1999Braun}. 
A much better approximation is given by the Density Matrix Renormalization Group \cite{1999Dukelsky}. 
But, the pairing Hamiltonian admits an exact solution worked out by Richardson at the beginning of the sixties \cite{1963Richardson,1964Richardson}. 
A more recent derivation of the exact solution can be found in ref. \cite{1999VonDelft}. 
The first application of the Richardson exact solution was done in ultra-small grains system 
\cite{1999Sierra,2000SierraDukelsky}. 
References \cite{1999Sierra,2000SierraDukelsky} and \cite{1999VonDelft} marks the resurgence of the Richardson's exact solution of the pairing Hamiltonian. 
The acknowledge to Richardson in refs. \cite{2000SierraDukelsky,1999VonDelft} constitutes a recognition to him after forty years in the oblivion.

The Richardson exact solution has been used to study the effect of the resonant single-particle states on the pairing Hamiltonian \cite{2003Hasegawa}. 
In ref. \cite{2006Pittel} the authors gave an interpretation of the pair energies from the Richardson solution. They relate the pairing correlations with the pair energies distribution in the complex plane. 
The pairing Hamiltonian is not only exactly soluble but also quantum integrable  \cite{1997Cambiaggio,2000Sierra,2001Amico,2001Dukelsky}. 
Besides the constant pairing, a very special kind of separable pairing interaction also admits an exact solution \cite{2007Balantekin,2011Dukelsky}. 
A review on exact solutions of the pairing Hamiltonian can be found in the ref. \cite{2004Dukelsky}. 

The pairing Hamiltonian approximates the influence of the  
residual interaction acting among the valence states lying close to the Fermi 
level.  However, it is an 
open question how one must treat pairing in the continuum. Previous studies 
on the contribution from the continuum to pairing have been reported in refs. 
\cite{2000Sandulescu,2001Kruppa,2003Hasegawa}. 

In this paper we reformulate 
the problem of determining the exact eigenenergies of the pairing Hamiltonian 
when the continuum is included. Real and complex energy representations of 
the continuum are used. The BCS approximation is not a convenient tool 
to treat many-body pairing close to the drip 
line \cite{1996Dobaczewsky,2012IdBetan}. It is the intention of this paper 
to give an exact treatment of the many-body pairing which overcomes the drawbacks of the BCS 
treatment.

The paper is organized as follows. Section \ref{sec.method} briefly reviews the 
derivation of the Richardson equations with the continuum represented on the 
real energy axis or in the complex energy plane.  In Sec. \ref{sec.result} the 
low lying states of even Carbon isotopes are evaluated and a comparison of
the solutions using the real energy representation are compared with  the ones 
obtained in the complex energy representation. The trajectory of the
pair energies are analyzed as a function of the pairing strength. The 
continuum pair energies are introduced in this section. Finally,  
Sec. \ref{sec.discussion} summarizes the main results of the paper.

\section{Method}\label{sec.method}
In this section the Richardson equations for a continuum basis is given. 
First the continuum is included by enclosing the system in a large spherical 
box.
After the final equations have been obtained, we take the limit of the box to 
infinity and introduce the single particle level density. 
In  order to avoid the Fermi gas we take the derivative of the phase shift 
for the continuum part of the single particle level density \cite{1937Beth}. 
Finally, we parametrized the CSPLD for the resonant partial waves and make the 
analytic continuation to the complex energy plane.

\subsection{System in a Box}
In this sub-section we follow the derivation of the exact solution as it was 
given by Jan Von Delft and Fabian Braun in ref. \cite{1999VonDelft}. The 
inclusion of the system in a large spherical box provides a finite discrete 
set of negative (bound) energies and an 
infinite discrete set of positive (continuum) energies. 
Let us called $\varepsilon_a$ the discrete energy with degeneracy $2j_a+1$, 
with $\alpha=\{ a, m_\alpha \}=\{ n_a, l_a, j_a, m_\alpha \}$. The pairing 
Hamiltonian is given by,
\begin{equation}
 H_P=\sum_{\alpha} \; \varepsilon_a \; c^\dagger_\alpha c_\alpha
   - G  \sum_{a m_\alpha>0} \; \sum_{b m_\beta>0}
             c^\dagger_\alpha c^\dagger_{\bar{\alpha}}  
             c_{\bar{\beta}}  c_\beta \;,
\end{equation}
with $c^\dagger_{\bar{\alpha}}=(-)^{j_a-m_\alpha} \; c^\dagger_{a -m_\alpha}$. We introduce the pair creation operator
\begin{equation}
  A^\dagger_a = \sum_{m_\alpha>0} \; c^\dagger_\alpha \; c^\dagger_{\bar{\alpha}} \;,
\end{equation}
which creates a pair of time reversal states with quantum number $a$.

Following Von Delft and Braun \cite{1999VonDelft}, who were inspired by a
suggestion by Richardson,  we propose the $N$-body ($N=2N_{\rm pair}$) 
eigenfunction as the antisymmetrised product of $N_{\rm pair}$ wave functions
as,
\begin{equation}
 |\Psi \rangle = \prod_{i=1}^{N_{\rm pair}} 
           \left( \sum_a \frac{A^\dagger_a}{2\varepsilon_a - E_{p_i}}  \right) |0\rangle \;,
\end{equation}
where the energies $E_{p_i}$ are related to the eigenvalues $E$ of the 
Hamiltonian $H_P$ by 
\begin{equation} \label{eq.e}
 E=\sum_{i=1}^{N_{\rm pair}} \; E_{p_i}\;.
\end{equation}

In order to meet the eigenvalue equation $H_P |\Psi \rangle = E \; |\Psi \rangle$, the parameters $E_{p_i}$, called pair energies, must verify the following set of $N_{\rm pair}$ couple system of equations \cite{1999VonDelft}
\begin{equation} \label{eq.rich}
 1 - \frac{G}{2} \sum_a \frac{2j_a+1}{2 \varepsilon_a - E_{p_i}} 
   + 2G \; \sum_{j \ne i}^{N_{\rm pair}} \; \frac{1}{E_{p_j} - E_{p_i}} = 0 \;,
\end{equation}
where the first summation contains negative and positive energies. The 
interpretation of this set of equations, called Richardson equations, is that 
the many-body fermions with pairing force behave like the many-boson system 
with one-body force. Both systems are described by the same wave function with 
the difference that the fermions have to satisfy the Richardson 
equations (\ref{eq.rich}) in order to fulfill the Pauli 
principle \cite{1964Richardson,1996VonDelft}.

\subsection{Continuum Real Energy}
In making the limit of the box to infinity the single particle states becomes 
more and more dense. In that limit the sum becomes an integral, i. e.,
\begin{equation}
 \sum_a \; (2j_a+1) \xrightarrow{V\rightarrow \infty} 
          \int_{-\infty}^\infty \; \tilde{g}(\varepsilon) \; d\varepsilon \;.
\end{equation}

The single particle density $\tilde{g}(\varepsilon)$ is the sum of the bound
(negative energy) states plus the continuum (positive energy) states.
We make the 
Anzatz that the single particle density in the continuum is given by the 
derivative of the phase shift \cite{1937Beth},
\begin{equation}
 \tilde{g}(\varepsilon) = \sum_b \; (2j_b + 1) \; \delta(\varepsilon-\varepsilon_b) +
                          \sum_c \; \frac{2j_c + 1}{\pi} \; \frac{d\delta_c}{d\varepsilon} \;,
\end{equation}
the index $b=(n_b,l_b,j_b)$ refers to bound states and $c=(l_c,j_c)$ to 
continuum states. The first summation is over the valence bound states while 
the second one is over the continuum partial waves. In practical applications
an upper limit $l_{\rm max}$ is set for the number of partial waves.

The Richardson equations in a representation which includes the continuum becomes,
\begin{eqnarray} \label{eq.rich_cont}
 & & 1 - \frac{G}{2} \sum_b \frac{d_b}{2 \varepsilon_b - E_{p_i}} 
   - \frac{G}{2} \int_0^\infty d\varepsilon  \; \frac{g(\varepsilon)}{2 \varepsilon - E_{p_i}} \nonumber \\ 
 & &   + 2 G \sum_{j \ne i} \; \frac{1}{E_{p_j} - E_{p_i}} = 0 \;,
\end{eqnarray}
where the factor $d_b=2j_b+1-2N_b$ takes into account the 
blocking effect of the $N_b$ unpaired states \cite{1964Richardson}. The
CSPLD becomes,
\begin{equation} \label{eq.g}
 g(\varepsilon) = \sum_c^{l_{\rm max}} \frac{2j_c+1}{\pi} \frac{d\delta_c}{d\varepsilon} \;.
\end{equation}

\subsection{Continuum Complex Energy}
The presence of the single particle resonances appear in the CSPLD, as well
as in the 
cross sections, as bumps. They correspond to states in the continuum (positive 
energy states) which are well localized inside the nuclear surface for a time 
greater 
than the characteristic nuclear time \cite{2001Cottingham}. One can thus
split the summation in resonant ($r$) and non-resonant ($nr$) (background) 
contributions as,
\begin{eqnarray}
 g(\varepsilon) &=& g_{_{\rm Res}}(\varepsilon) + g_{_{\rm Bckg}}(\varepsilon)  \;, \\
 g_{_{\rm Res}}(\varepsilon) &=& \sum_r \frac{2j_r+1}{\pi} \frac{d\delta_r}{d\varepsilon} \;, \\
 g_{_{\rm Bckg}}(\varepsilon) &=& 
     \sum_{nr} \frac{2j_{nr}+1}{\pi} \frac{d\delta_{nr}}{d\varepsilon} \label{eq.gbckg} \;,
\end{eqnarray}

The single particle density for the resonant states at energies $\epsilon_r$ 
and widths $\Gamma_r$ can be written as \cite{1988Kukulin}.
\begin{equation}
 g_{_{\rm Res}}(\varepsilon) = \sum_r \frac{2j_r+1}{\pi} \frac{\Gamma_r/2}{(\varepsilon - \epsilon_r)^2+(\Gamma_r/2)^2} \;.
\end{equation}

The resonant parameters can be represented by a single complex 
number $\varepsilon_r = \epsilon_r - i \; \Gamma_r/2$ which corresponds to 
the eigenvalue of the mean-field Hamiltonian with pure outgoing boundary 
condition \cite{1968Berggren}. By rotating the integration contour of the 
resonant part of the CSPLD to the negative imaginary axis, and applying the 
Cauchy theorem, one gets the Richardson equations in terms of the 
complex energy states,
\begin{widetext}
\begin{equation}\label{eq.rich2}
  1 - \frac{G}{2} \sum_b \frac{d_b}{2 \varepsilon_b - E_{p_k}} 
   - \frac{G}{2} \sum_r \frac{2j_r+1}{2 \varepsilon_r - E_{p_k}} 
   - \frac{G}{2} \int_0^\infty d\varepsilon  \; \frac{g_{_{\rm Bckg}}(\varepsilon)}{2 \varepsilon - E_{p_k}} 
   - \frac{G}{2} \int_0^\infty d\varepsilon  \; \frac{g_{_{\rm CxBckg}}(\varepsilon)}{2 \varepsilon - iE_{p_k}} 
     + 2 G \sum_{l \ne k} \; \frac{1}{E_{p_l} - E_{p_k}} = 0 \;,
\end{equation}
\end{widetext}
where
\begin{equation}
 g_{_{\rm CxBckg}}(\varepsilon)= - \sum_r \frac{2j_r+1}{\pi} 
     \frac{\Gamma_r/2}{(\varepsilon - i \epsilon_r)^2 - (\Gamma_r/2)^2}  \;.
\end{equation}

In an overstatement (the ``density'' $g_{_{\rm CxBckg}}$ can not  be 
defined outside the integral) one could say that the background contribution to 
the Richardson equation has a real part coming from the non-resonant scattering 
partial wave states $g_{_{\rm Bckg}}$ and a complex contribution $g_{_{\rm CxBckg}}$ 
which is a remnant of the complex analytic extension from $g_{_{\rm Res}}$. Because 
the presence of the complex energy Gamow states in the second summation in Eq. 
(\ref{eq.rich2}), the complex contribution of $g_{_{\rm CxBckg}}$ is necessary to 
make $E=\sum_i \; E_{p_i}$ real. In Eq. (\ref{eq.rich2}) we have assumed that 
there is not blocking effect due to continuum states. 

For the seniority zero case and neglecting the background, Eq. (\ref{eq.rich2}) 
reduces to the Richardson equations in the Gamow basis introduced in 
ref. \cite{2003Hasegawa}. In this case the complex pairing energies are not 
complex conjugate to each other, i.e. $E=\sum_i \; E_{p_i}$ may be complex.

\subsection{Exact Spectrum}
The solution of the Richardson equations (\ref{eq.rich_cont}) with 
the ``boundary condition'',
\begin{equation} \label{eq.limit}
 lim_{G\rightarrow 0^+} E_{p_i}=2\varepsilon_{p_i} \;,
\end{equation}
and the blocking effect, determine the ground state and the excited state 
energies of the pairing Hamiltonian.

The $^{12}$C nucleus has three bound configurations (sec. \ref{sec.sp}).  
The first $(1)$ and second $(2)$ 
configurations can accommodate a single pair, while the third 
configuration $(3)$ can accommodate three pairs. The configurations 
$(1)$, $(2)$, and $(3)$ are related to the single particle states $0p_{1/2}$, 
$1s_{1/2}$, and $0d_{5/2}$, respectively. 
Then $\varepsilon_{p_1}=\varepsilon_{0p_{1/2}}$, $\varepsilon_{p_2}
=\varepsilon_{0s_{1/2}}$, and $\varepsilon_{p_3}=\varepsilon_{p_4}
=\varepsilon_{p_5}=\varepsilon_{0d_{5/2}}$. From the bound configurations we 
can accommodate up to five pairs ($^{22}$C). Because the inclusion of the 
continuum we will be able to go beyond the nucleus $^{22}$C.

\subsubsection{Ground State}
The ground state (g.s.) configuration for a system with $N_{\rm pair}$ corresponds 
to fill the lowest $N_{\rm pair}$ configurations by solving the Richardson 
Eq. (\ref{eq.rich_cont}) with the blocking coefficient $d_b=2j_b+1$ because 
the g.s. has seniority zero and there are no unpaired states (all $N_b=0$). For 
example, the g.s. of the isotope $^{14}$C corresponds to solving one single 
Richardson equation (\ref{eq.rich_cont}) with the boundary condition $lim_{G\rightarrow 0^+} 
E_{p_1}=2\varepsilon_{p_1}$. Let us called this configuration $(1)^2$. 
The g.s. of 
the isotope $^{16}$C corresponds to solving two Richardson equations (\ref{eq.rich_cont}) with the 
boundary conditions $lim_{G\rightarrow 0^+} E_{p_1}=2\varepsilon_{p_1}$ and
 $lim_{G\rightarrow 0^+} E_{p_2}=2\varepsilon_{p_2}$. This is the 
configuration $(1)^2(2)^2$, and so on. The ground state energy $E$ is given by 
Eq. (\ref{eq.e}) with $N_{\rm pair}=1,2$ and so on.

\subsubsection{Excited States}
We have to distinguish between excited states with seniority zero and seniority two. 
\\
\textit{Seniority Zero ($\nu=0$):} The seniority zero excited states are found by solving 
as many equations (\ref{eq.rich_cont}) as pairs, like for the g.s., but with a boundary condition 
other than the ground state. For example, the first and second $0^+$ excited 
states of $^{14}$C are found as the solution of a single equation with the 
boundary conditions $lim_{G\rightarrow 0^+} E_{p_2}=2\varepsilon_{p_2}$, 
and $lim_{G\rightarrow 0^+} E_{p_3}=2\varepsilon_{p_3}$, respectively. We 
called such configurations $(2)^2$ and $(3)^2$. As a second example let us 
consider the first $0^+$ excited state of $^{18}$C. It is found by solving 
three equations (\ref{eq.rich_cont}) with the boundary conditions $lim_{G\rightarrow 0^+} 
E_{p_1}=2\varepsilon_{p_1}$, $lim_{G\rightarrow 0^+} 
E_{p_2}=2\varepsilon_{p_3}$, and $lim_{G\rightarrow 0^+} E_{p_3}=2\varepsilon_{p_3}$. We called  this configuration $(1)^2(3)^4$. The energy $E$
of the $\nu=0$ excited state is like Eq. (\ref{eq.e}) but using the excited 
pair energies.

\textit{Seniority Two ($\nu=2$):} The seniority two states are found by 
solving $N_{\rm pair}=(A-12)-\nu$ equations (\ref{eq.rich_cont}), where $A$ is the mass number of the 
isotope. This is one equation less than the number of pairs in the ground state. 
The factor $d_b$ in Eq. (\ref{eq.rich_cont}) is given by $d_b=2j_b+1-2N_b$, where $b$ labels the blocking configuration. 
For example, to find the $\nu=2$ states in $^{14}$C one does not need to solve any equation
since $N_{\rm pair}=(14-12)-2=0$. The $\nu=2$ state energy is just the sum of the single 
particle energies $E=\varepsilon_l + \varepsilon_m$ of the
unpaired levels $l$ and $m$. Let us assumed that the blocking states for the isotope $^{16}$C are the 
configurations $(2)$ and $(3)$, i. e. $N_1=0$, and $N_2=N_3=1$. Then, we have 
to solve a single equation with $d_1=2$, $d_2=0$ and $d_3=8$ and the boundary 
condition $lim_{G\rightarrow 0^+} E_{p_1}=2\varepsilon_{p_1}$. Let us call 
this configuration $(1)^2(2)(3)$ which gives the degenerate levels $2^+,3^+$. The energy of such a state is 
$E=E_{p_1} + \varepsilon_2 + \varepsilon_3$. As the last example, let us 
consider the first $2^+,4^+$ states in $^{18}$C. This level in found by solving 
two equations with the boundary conditions $lim_{G\rightarrow 0^+} E_{p_1}=
2\varepsilon_{p_1}$ and $lim_{G\rightarrow 0^+} E_{p_2}=2\varepsilon_{p_2}$, 
and with $N_1=N_2=0$ and $N_3=2$. The energy of this last state is $E=E_{p_1} 
+ E_{p_2} + 2\varepsilon_3$.

\subsection{Determination of the Pairing Strength} \label{sec.ps}
In order to determine the strength $G$ we consider the neutron pairing energy 
$P_{\rm Exp}(2N_{\rm pair})$ for a system of $N=2N_{\rm pair}$ valence neutrons 
\cite{2007Suhonen}
\begin{equation}
 P_{\rm Exp}(2N_{\rm pair}) = 2E(2N_{\rm pair}-1) - E(2N_{\rm pair}) -E(2N_{\rm pair}-2) \;.
\end{equation}

The pairing energy in the Richardson model is related to the last pair 
energy $E_{p_{N_{\rm pair}}}$ as follows \cite{1964Richardson},
\begin{equation}
 P_{\rm Rich} = 2 \varepsilon_{p_{N_{\rm pair}}} - Re\left[ E_{p_{N_{\rm pair}}}(2N_{\rm pair}) \right] \;.
\end{equation}

By imposing the condition $P_{\rm Exp}=P_{\rm Rich}$ one finds the strength $G$ which 
reproduces $E_{p_{N_{\rm pair}}}$.

\subsection{Determination of the Resonant Partial Waves} \label{sec.tau}
The criterion to decide whether a given partial wave is resonant is to search
for the poles $\varepsilon_{lj}=\epsilon_{lj} - i \frac{\Gamma_{lj}}{2}$ of 
$S_{lj}$. A physical resonance should satisfy that the half-life calculated 
with the 
imaginary part $\Gamma_{lj}/2$ of the pole $\tau=\frac{\hbar \; ln\;2}{\Gamma_{lj}}$ 
is bigger than the characteristic time $\tau_c=2.6\times 10^{-23} 
\times A^{1/3}$ sec. \cite{2001Cottingham}. The physical meaning of 
this criterion is that the particle has enough time to interact with the 
system before it decays.

\section{Applications} \label{sec.result}
\subsection{Parameters}
This sub-section aims to define the real and complex single particle representations. 
The parameters for the interaction are also set up here. 
The real energy representation consists of a finite discrete set of bound states plus a positive real continuum set of scattering states. 
While the complex energy representation consists of a finite discrete set of bound and Gamow states plus a complex continuum set of ``scattering states". 
In the complex energy representation we named resonant continuum the set of Gamow states and non-resonant continuum to the scattering states with complex energy.

\subsubsection{Single Particle Representation} \label{sec.sp}
\begin{figure}[ht]
\vspace{6mm}
\includegraphics[width=0.45\textwidth]{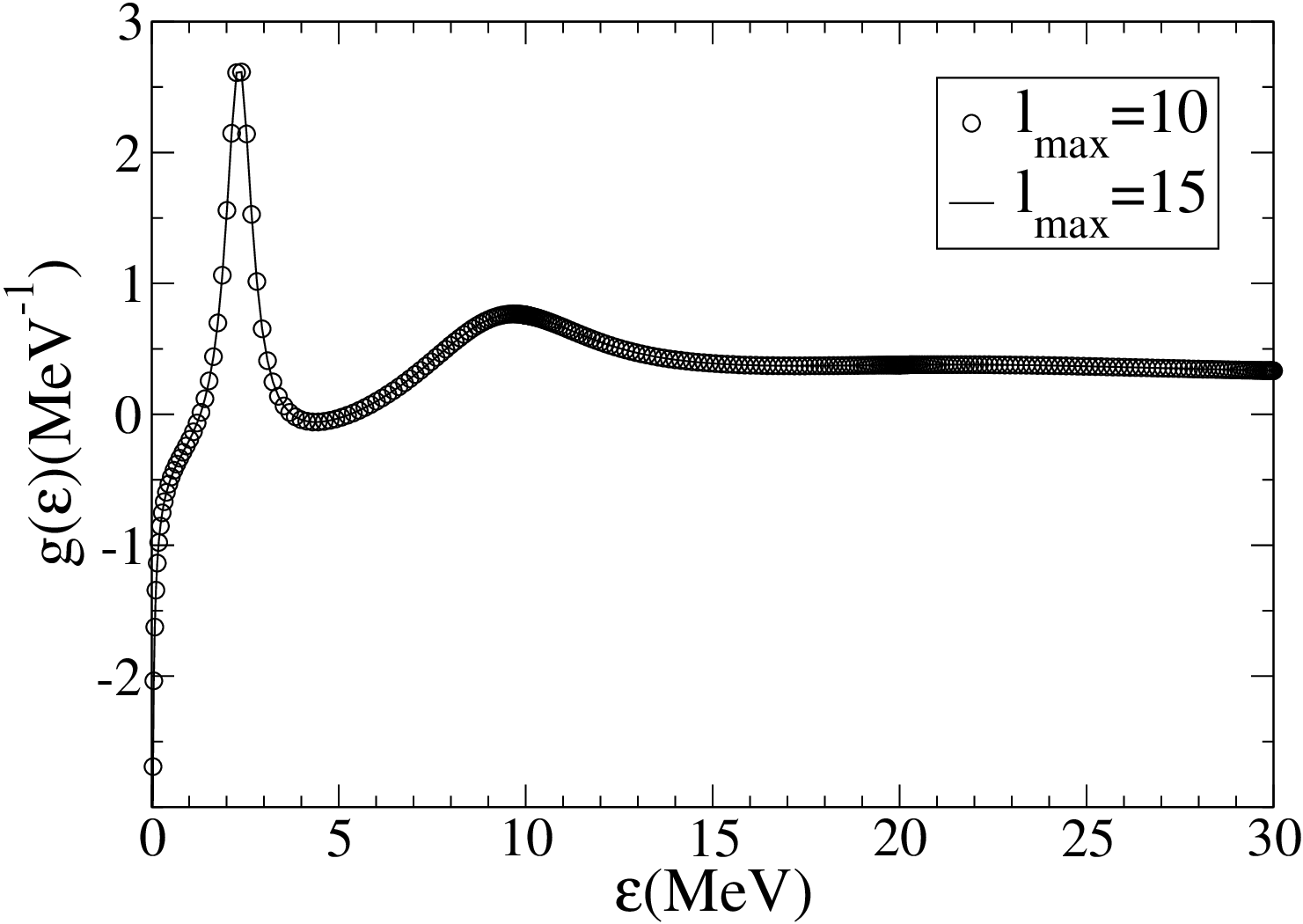}
\caption{\label{fig.1} Neutron CSPLD in $^{12}$C for two different angular 
momentum cutoff.}
\end{figure}

The experimental single particle energies in $^{13}$C were taken from ref. \cite{nndc}: $\varepsilon_{0p_{1/2}}=-4.946$ MeV, $\varepsilon_{1s_{1/2}}=-1.857$ MeV, and $\varepsilon_{0d_{5/2}}=-1.093$ MeV. The single particle density of $^{13}$C was calculated with the program \cite{1995Ixaru} with the following Woods-Saxon parameters: $V_0=55.1$ MeV, $V_{so}=10.5$ MeV, $a=a_{so}=0.7$ fm, $r_0=r_{so}=1.27$ fm. Fig. \ref{fig.1} compares the CSPLD for $l_{\rm max}=10$ and $l_{\rm max}=15$. It shows 
that a cut-off of $l=10$ in Eq. (\ref{eq.g}) is enough for this system.

The negative contribution in Fig. \ref{fig.1} is due to the dominance of 
the $s_{1/2}$ state at low energy. In accordance to the Levinson theorem, it 
must be a negative contribution for each bound state. For the $s_{1/2}$ state 
this negative contribution is close to the continuum threshold. The resonant 
behavior around $2$ MeV is due to the resonant state $d_{3/2}$, while the 
one around $10$ MeV is due the wide resonance $f_{7/2}$. Using the code Gamow \cite{1995Ixaru} 
we find the following energies for these two states, $\varepsilon_{0d_{3/2}}
=(2.2671;-0.416)$ MeV, and $\varepsilon_{0f_{7/2}}=(9.288;-3.040)$ MeV.

\subsubsection{Pairing Strength}
From the experimental mass excess table we got for the pairing energy 
$P_{\rm Exp}$ of the isotope $^{14}$C, $P_{\rm Exp}(^{14}C)=1.516$ MeV. In the 
Richardson model the pairing energy is related to the pair energy through 
$P(2N_{\rm pair})=2\varepsilon_{p_{N_{\rm pair}}} - E_{p_{N_{\rm pair}}}$ (Sec. 
\ref{sec.ps}). For $^{14}$C, $N_{\rm pair}=1$, then $P(^{14}C)=2\varepsilon_{p_1} 
- E_{p_1}$ with $2\varepsilon_{p_1}=-9.989$ MeV and $E_{p_1}=-11.408$ MeV. 
In order to reproduce $E_{p_1}$ with a cutoff energy at $30$ MeV, one must 
take $G=0.7786$ MeV. Using the parametrization $G=\frac{\chi}{A}$ we 
obtained $\chi=10.900$ for $A=14$. This value of $\chi$ is used for all 
Carbon isotopes. Table \ref{table.g} lists the value of the pairing strength 
for each Carbon isotope.
\begin{table}[ht]
\caption{\label{table.g}Pairing strength used for the Carbon isotopes.}
\begin{ruledtabular}
\begin{tabular}{cc}
 Isotope  & $G$[MeV] \\
\hline
 $^{14}$C & $0.7786$ \\
 $^{16}$C & $0.6813$ \\
 $^{18}$C & $0.6056$ \\
 $^{20}$C & $0.5450$ \\
 $^{22}$C & $0.4955$\\
 $^{24}$C & $0.4542$
\end{tabular}
\end{ruledtabular}
\end{table}

\subsection{Results: Real Energy Representation} \label{sec.real}
After the model space and the interaction are set up one can evaluate
physical magnitudes. In this subsection we are going to 
calculate the ground state energy of the carbon isotopes $^{14}$C to 
$^{24}$C and the low energy spectrum of the isotopes $^{14}$C to $^{20}$C.

\subsubsection{Ground-state Energy}
\begin{figure}[ht]
\vspace{6mm}
\includegraphics[width=0.45\textwidth]{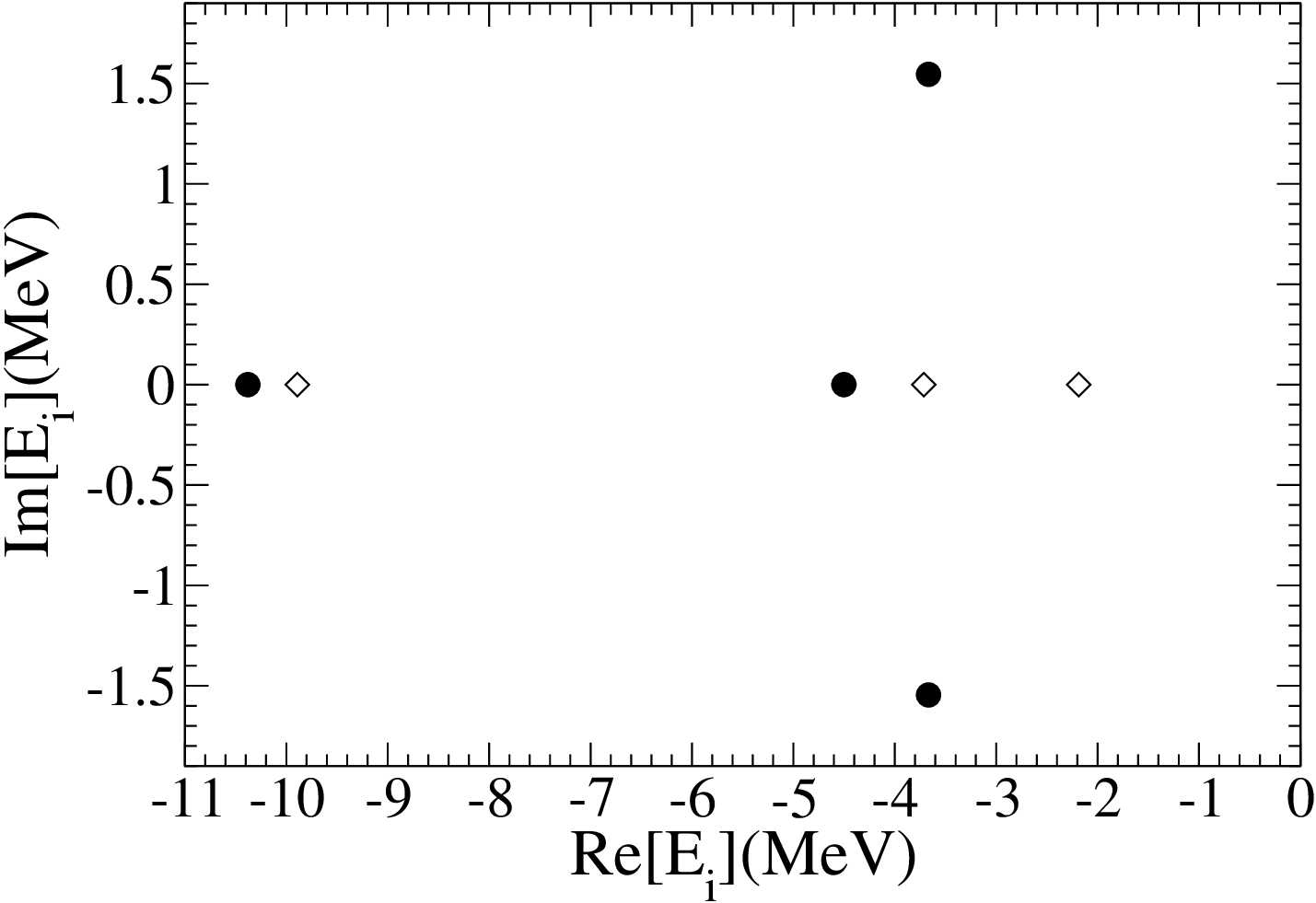}
\caption{\label{fig.20C} Distribution of the four pair energies in $^{20}$C 
isotope (dark dots). The white diamond correspond to the pair energies for 
$G=0$, i.e. $E_i=2\varepsilon_i$.}
\end{figure}

\begin{figure}[ht]
\vspace{6mm}
\includegraphics[width=0.45\textwidth]{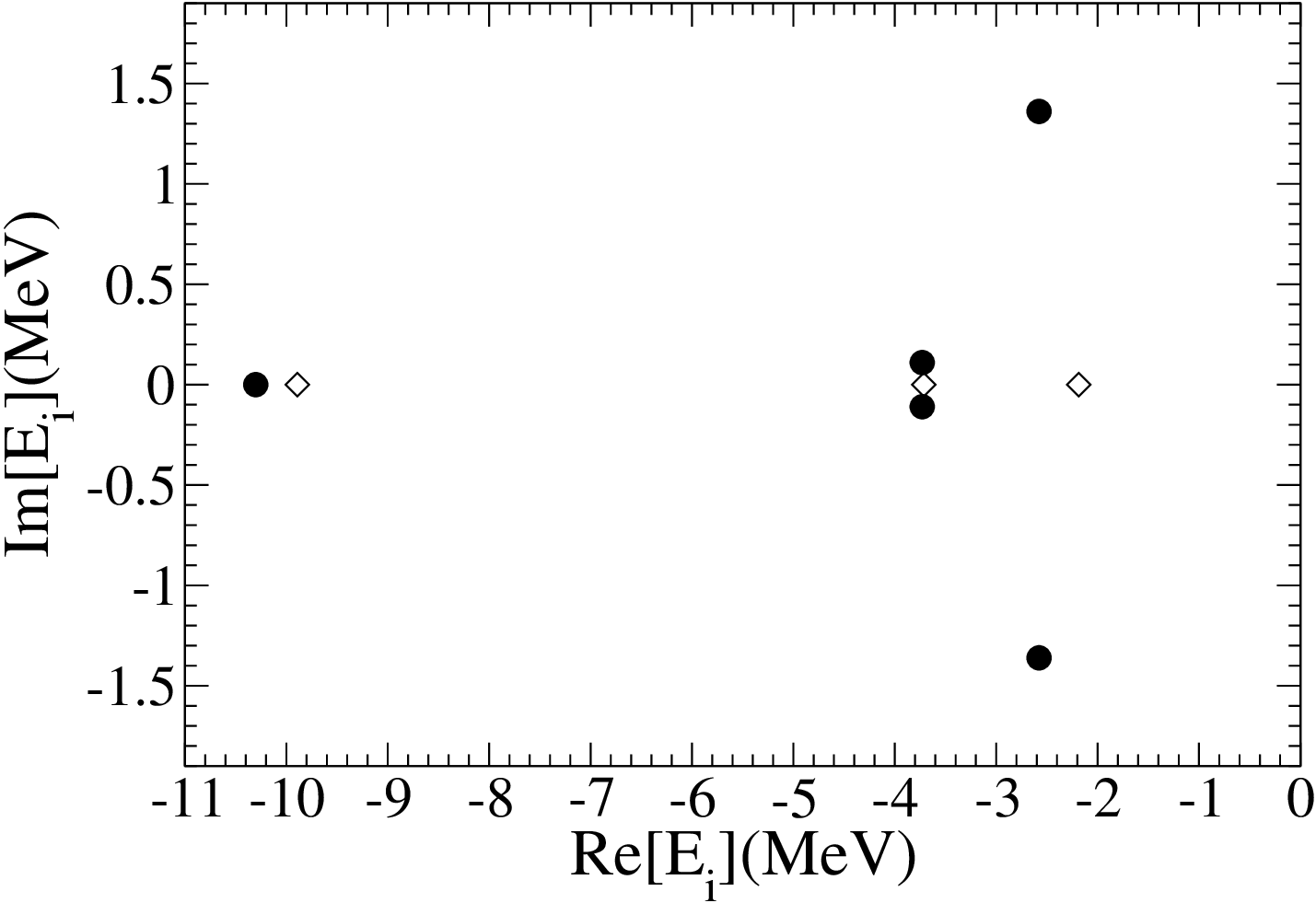}
\caption{\label{fig.22C} Like fig. \ref{fig.20C} for the five pair energies 
in $^{22}$C.}
\end{figure}

\begin{figure}[ht]
\vspace{6mm}
\includegraphics[width=0.45\textwidth]{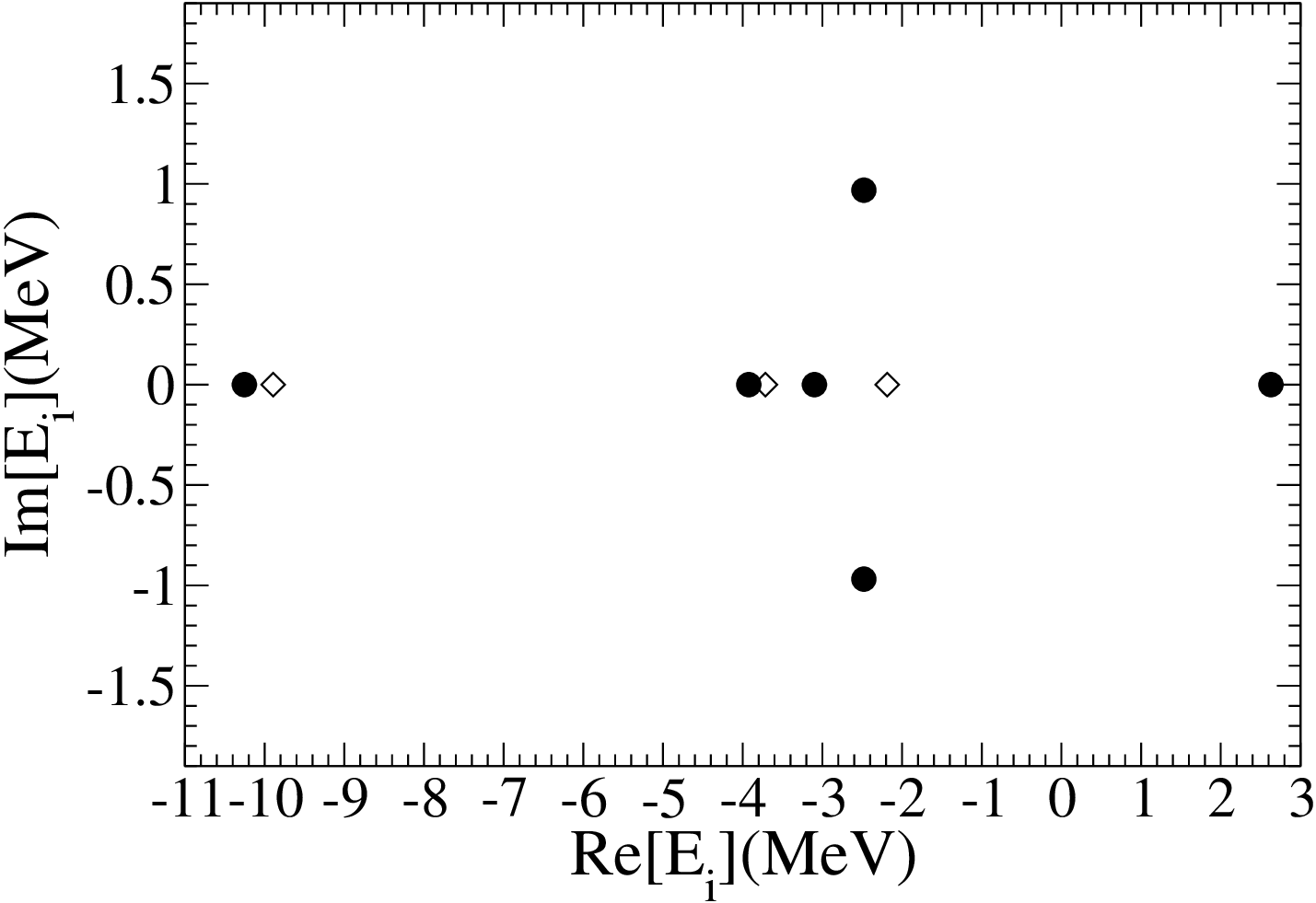}
\caption{\label{fig.24C} Like fig. \ref{fig.20C} for the six pair energies in
$^{24}$C.}
\end{figure}

Solving the Richardson equations (\ref{eq.rich_cont}) for the ground state of each carbon isotope, 
we obtained a set of pair energies $E_i$ (we set $E_i$ for $E_{p_i}$) as it is 
shown in table \ref{table.epair}. Complex pair energies appear in complex 
conjugate pairs to give a real eigenenergy. The distribution of the pair 
energies gives information about the structure of the 
many-body wave function. 
As the many-body state becomes more collective, more pairs accommodate themselves in a 
parabola-like distribution \cite{2006Pittel}. Let us quantized roughly the 
degree of collectivity $\gamma$ as the 
ratio of the number of pairs which participate in a parabola versus the total 
number of pairs. We will do this  
for system with at least four pairs. We observe a high degree of collectivity 
as one approaches 
the threshold, while the collectivity abruptly drops in the continuum. 
Figures \ref{fig.20C}, \ref{fig.22C} and \ref{fig.24C} show the distribution of 
the pair energies in the complex energy plane for the isotopes $^{20}$C, 
$^{22}$C and $^{24}$C, respectively. 

\begin{table}[ht]
\caption{\label{table.epair}Pair energies $E_i$ and ground state energies 
$E_0$ relative to carbon $^{12}$C for the Carbon isotopes $^{14}C-^{24}$C. We 
used $E_i$ for $E_{p_i}$. The collectivity parameter $\gamma$ was defined in 
the text.}
\begin{ruledtabular}
\begin{tabular}{ccllc}
 Isotope  & $N_{\rm pair}$ & $E_i$[MeV]                   & $E_0$[MeV] & $\gamma$ \\
\hline
 $^{14}$C & 1          & $E_1=-11.398$                & $-11.398$  & - \\
 $^{16}$C & 2          & $E_1=-10.681$                & $-17.051$  & - \\
          &            & $E_2=-6.370$                 &   & \\
 $^{18}$C & 3          & $E_1=-10.495$                & $-20.394$  & - \\
          &            & $E_{2,3}=(-4.950;\pm 1.262)$ &   & \\
 $^{20}$C & 4          & $E_1=-10.379$                & $-22.194$  & 0.75 \\
          &            & $E_2=-4.502$                 &   & \\
          &            & $E_{3,4}=(-3.667;\pm 1.546)$ &   & \\
 $^{22}$C & 5          & $E_1=-10.302$                & $-22.915$  & 0.8 \\
          &            & $E_{2,3}=(-3.729;\pm 0.110)$ &   & \\
          &            & $E_{4,5}=(-2.578;\pm 1.361)$ &   & \\
 $^{24}$C & 6          & $E_1=-10.254$                & $-19.605$  & 0.5 \\
          &            & $E_2=-3.924$                 &   & \\
          &            & $E_3=-3.099$                 &   & \\
          &            & $E_{4,5}=(-2.479;\pm 0.969)$ &   & \\
          &            & $E_6=2.630$                 &   & 
\end{tabular}
\end{ruledtabular}
\end{table}
 
\begin{figure}[ht]
\vspace{6mm}
\includegraphics[width=0.45\textwidth]{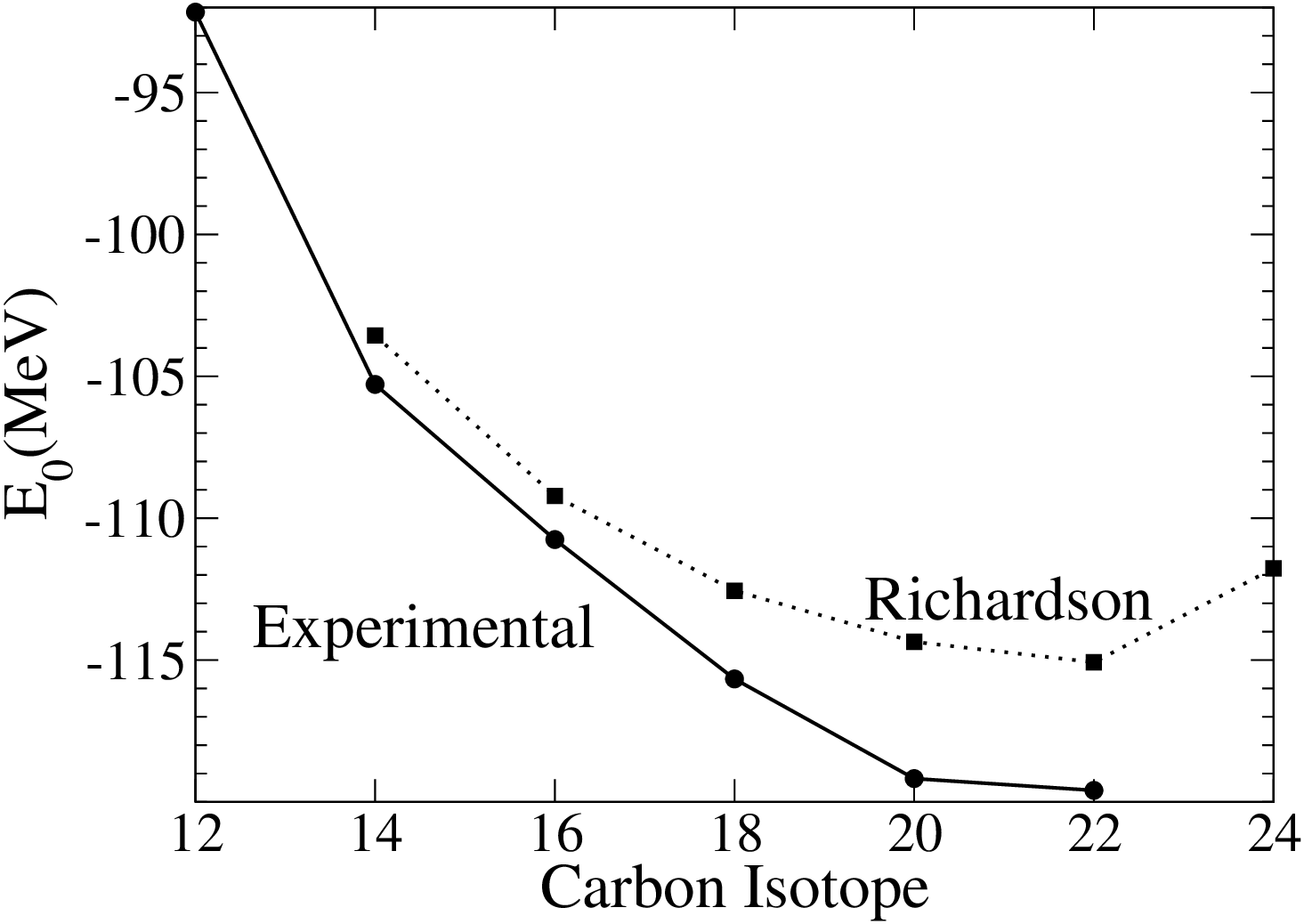}
\caption{\label{fig.be} Carbon isotopes ground-state energy.}
\end{figure}

Table \ref{table.epair} also shows the ground state energy $E_0$ of the Carbon 
isotopes $^{14}$C to $^{24}$C. 
Fig. \ref{fig.be} compares the calculated ground-state energy with the experimental one \cite{2003Audi}. 
It is found that the exact solutions follow the overall trend, i.e. the binding energy decreases faster at the beginning of the chain and decelerates when it approaches the drip line. The agreement with data worsen as the number of neutrons increases. 
Even when the pairing interaction is a schematic one, and not realistic, this investigation suggests that the nucleus $^{24}$C is unbound.

\subsubsection{Carbon Isotopes Spectrum}
It is worthwhile to 
compare the experimental spectrum with the 
exact solutions of the schematic pairing Hamiltonian corresponding to 
the cases of seniority-zero and seniority-two.

\begin{figure}
\vspace{6mm}
\includegraphics[width=0.45\textwidth]{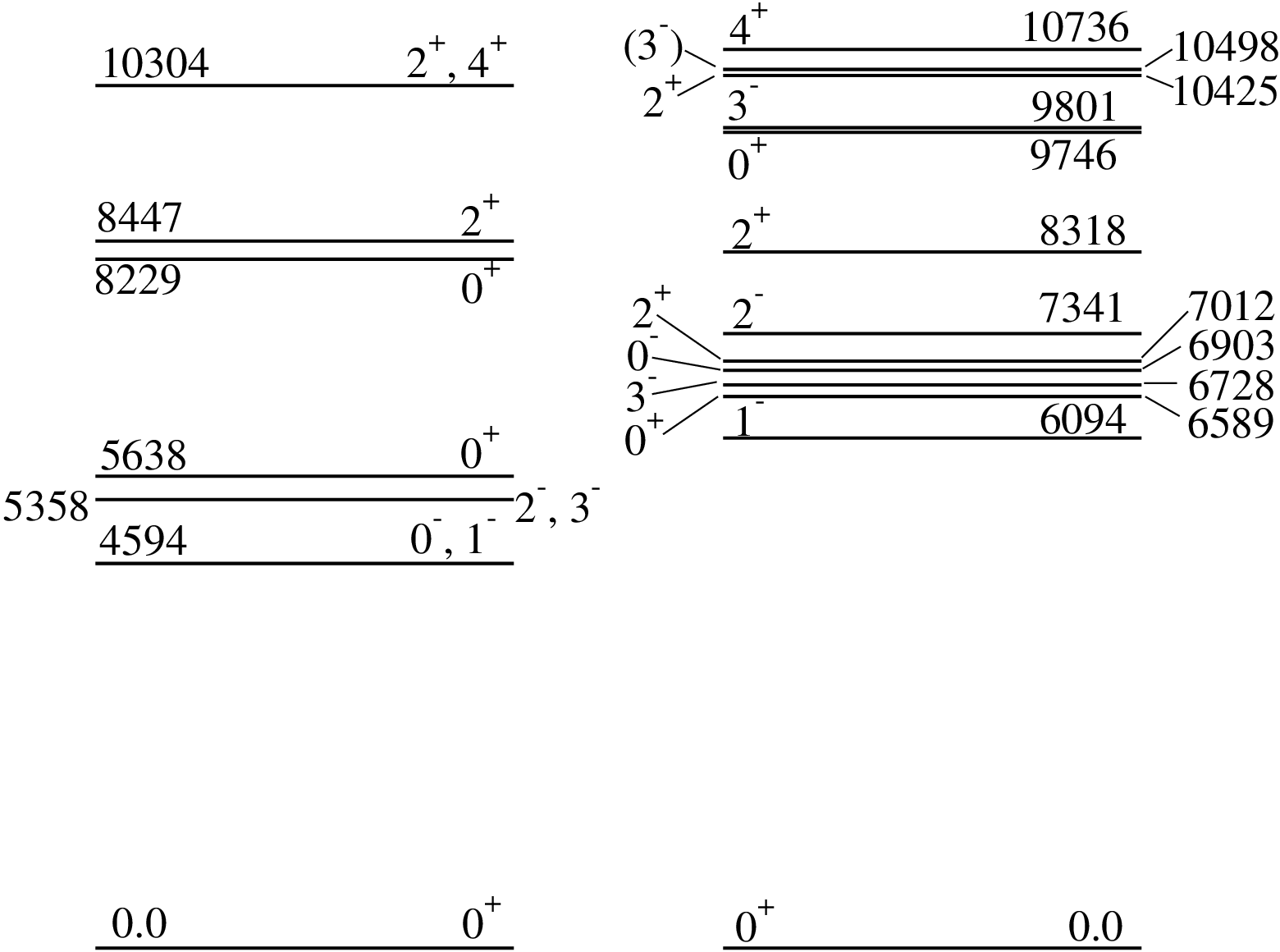}
\caption{\label{fig.c14} Exact low energy spectrum of $^{14}$C for seniority 
zero and two compared with experimental levels \cite{2003Audi}. The energies 
are in keV.}
\end{figure}

\textbf{$^{14}$C Spectrum:}
Table \ref{table.14C} gives the excitation spectrum (last column) with 
respect to the ground state configuration $(1)^2$. The seniority $\nu$, the 
pair energies and the number of pair $N_{\rm pair}=(A-12)-\nu$ ($A$ the mass 
number) are also given. 
Figure \ref{fig.c14} compares the calculated levels from table 
\ref{table.14C} with that of the experimental one. The quantum number of the 
first excited state $1^-$ is correctly found with $1.5$ MeV less energy. The 
$0^+_2$ and $0^+_3$ excited states are underestimated with respect to the 
experimental one by $0.951$ MeV and $1.517$ MeV respectively. The state $3^-$ 
is found $1.37$ MeV below the experimental one. The splitting between the 
states $3^-$ and $0^+_2$ is well reproduced: $280$  keV versus the experimental 
$175$ keV but in inverse order. We missed the first $2^+$ state and found 
a $2^+$ at only $129$ keV from the second experimental $2^+$. The near 
degenerate experimental states $2^+$ and $4^+$ around $10$ MeV are well reproduced.

\begin{table}
\caption{\label{table.14C} Excited and pair energies of $^{14}$C. The energies 
are in MeV.}
\begin{ruledtabular}
\begin{tabular}{ccccccc}
 Config   & $\nu$ & State     & $N_{\rm pair}$ & $E_{p_i}$         & $E$       & $Ex$     \\
\hline
 $(1)^2$  & 0     & $0^+$     & 1          & $E_{p_1}=-11.398$ & $-11.398$ & $0$      \\
 $(1)(2)$ & 2     & $0^-,1^-$ & 0          &                   & $-6.803$  & $4.594$ \\
 $(1)(3)$ & 2     & $2^-,3^-$ & 0          &                   & $-6.039$  & $5.358$ \\
 $(2)^2$  & 0     & $0^+$     & 1          & $E_{p_2}=-5.760$  & $-5.760$  & $5.638$ \\
 $(3)^2$  & 0     & $0^+$     & 1          & $E_{p_3}=-3.168$  & $-3.168$  & $8.229$ \\
 $(2)(3)$ & 2     & $2^+$     & 0          &                   & $-2.950$  & $8.447$ \\
 $(3)(3)$ & 2     & $2^+,4^+$ & 0          &                   & $-1.093$  & $10.304$
\end{tabular}
\end{ruledtabular}
\end{table}

\begin{figure}
\vspace{6mm}
\includegraphics[width=0.45\textwidth]{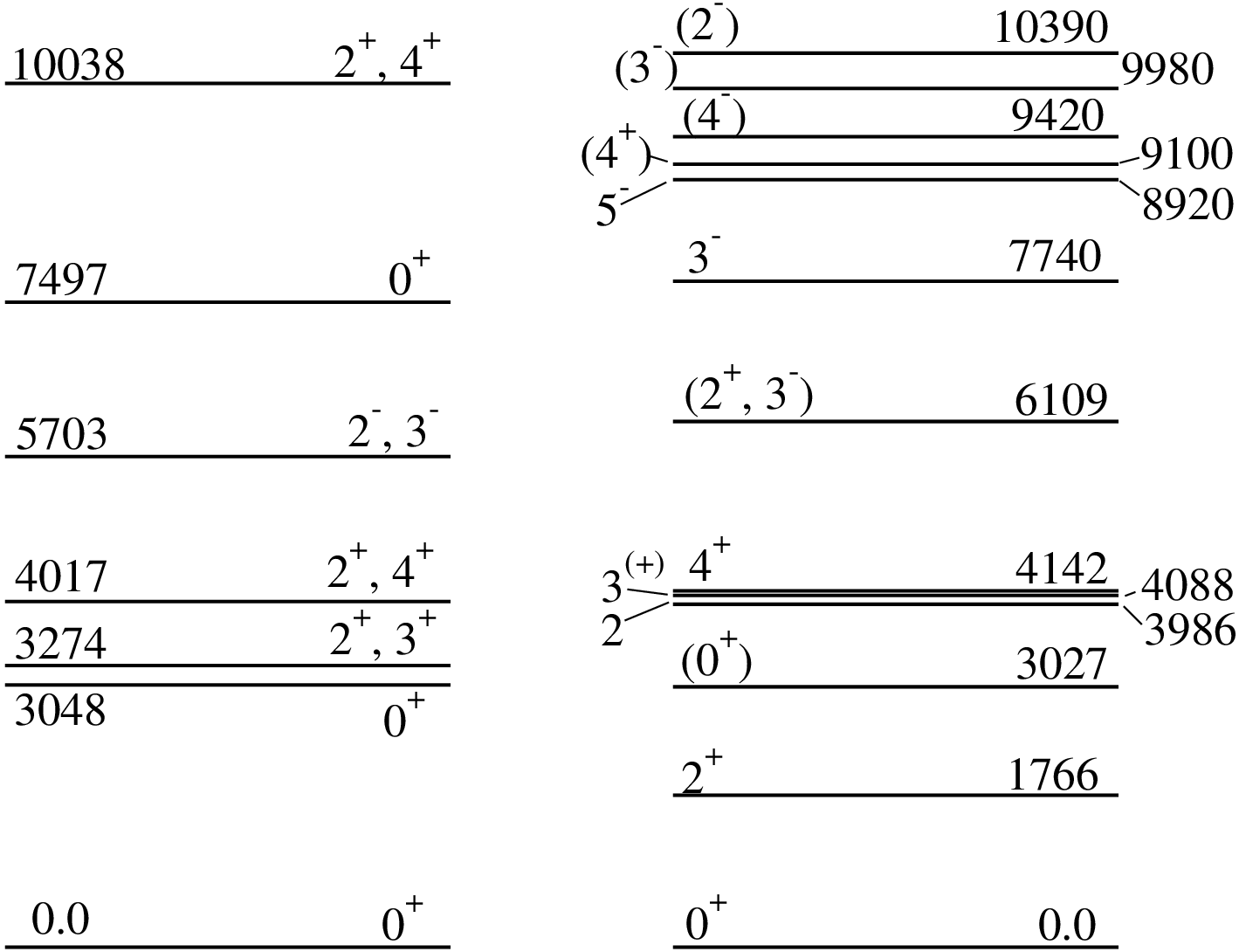}
\caption{\label{fig.c16} Like fig. \ref{fig.c14} for $^{16}$C. The 
experimental levels above $6.11$ MeV are from \cite{2003Bohlen}. The energies 
are in keV.}
\end{figure}

\textbf{$^{16}$C Spectrum:}
Table \ref{table.16C} shows the pair energies and the excitation spectrum with 
respect to the ground state configuration $(1)^2(2)^2$.
Fig. \ref{fig.c16} compares the calculated versus the experimental spectrum 
of $^{16}$C. The first excited $2^+$ state does not appear in our spectrum. The 
first $0^+$ excited state is very well reproduce with a difference of only $21$ 
keV. We found a $2^+$ state at $3.274$ MeV which may correspond to the 
experimental $2$ state at $3.986$ MeV. The first $4^+$ excited state is found 
only $125$ keV below the experimental one. The experimental $(3^-)$ is $406$ 
keV from the $3^-$ calculated state. In the exact spectrum appears a 
third $0^+$ state which does not appear in the experimental spectrum. Finally, 
the expaerimental $(4^+)$ state is $938$ keV from the $4^+$ calculated one. 
Summing up what we found for the nucleus $^{16}$C, the first $0^+$, $4^+$ 
and $3^-$ are reasonable well 
described by the pairing interaction.

\begin{table*}
\caption{\label{table.16C} Like table \ref{table.14C} for $^{16}$C.}
\begin{ruledtabular}
\begin{tabular}{ccccccc}
 Config        & $\nu$ & State     & $N_{\rm pair}$ & $E_i[MeV]$                        & $E[MeV]$       & $Ex[MeV]$     \\
\hline
 $(1)^2(2)^2$  & 0     & $0^+$     & 2          & $E_1=-10.681$                & $-17.051$ & $0$      \\
               &       &           &            & $E_2=-6.370$                 &           &          \\
 $(1)^2(3)^2$  & 0     & $0^+$     & 2          & $E_1=-10.823$                & $-14.003$ & $3.048$  \\
               &       &           &            & $E_3=-3.180$                 &           &          \\
 $(1)^2(2)(3)$ & 2     & $2^+,3^+$ & 1          & $E_1=-10.827$                & $-13.777$ & $3.274$ \\
 $(1)^2(3)(3)$ & 2     & $2^+,4^+$ & 1          & $E_1=-10.848$                & $-13.034$ & $4.017$ \\
 $(2)^2(1)(3)$ & 2     & $2^-,3^-$ & 1          & $E_2=-5.309$                 & $-11.348$ & $5.703$\\
 $(2)^2(3)^2$  & 0     & $0^+$     & 2          & $E_{2,3}=(-4.777,\pm 1.079)$ & $-9.554$  & $7.497$\\
 $(2)^2(3)(3)$ & 2     & $2^+,4^+$ & 1          & $E_2=-4.827$                 & $-7.013$  & $10.038$\\
\end{tabular}
\end{ruledtabular}
\end{table*}

\begin{figure}
\vspace{6mm}
\includegraphics[width=0.45\textwidth]{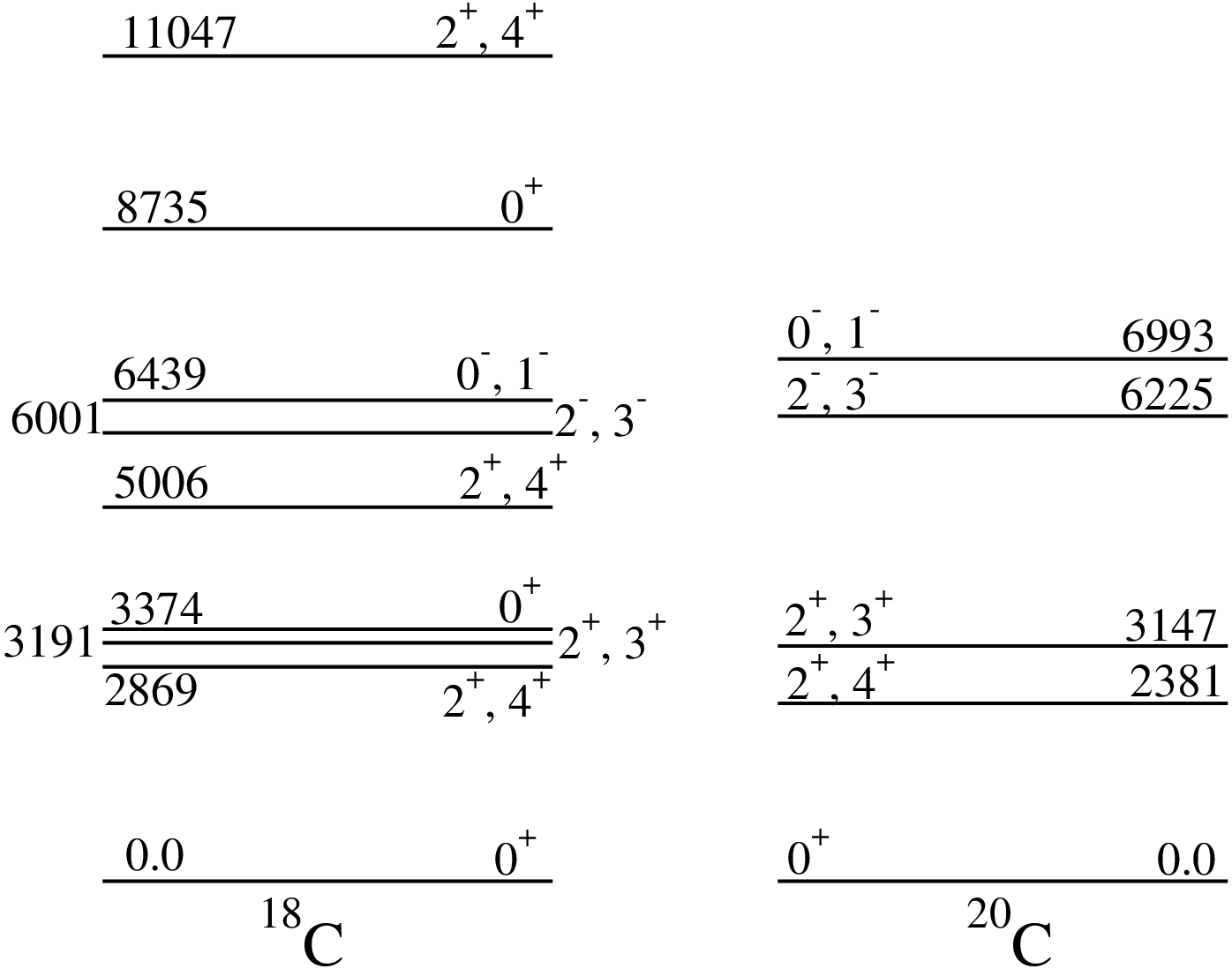}
\caption{\label{fig.c18} Exact low energy spectra of $^{18}$C and $^{20}$C 
for seniority zero and two. The energies are in keV.}
\end{figure}

\textbf{$^{18}$C and $^{20}$C Spectra:}
Tables \ref{table.18C} and \ref{table.20C} show the pair energies and the 
excitation spectrum with respect to the ground state configuration  for the 
three and four pair systems $^{18}$C and $^{20}$C respectively.
Figure \ref{fig.c18} shows the calculated exact eigenvalue of the pairing 
Hamiltonian for $^{18}$C and $^{20}$C. Experimentally, only one excited state 
in $^{18}$C is known. It is a $(2^+)$ state at $1620$ keV from the $(0^+)$ 
ground state. Considering what we learn in the previous spectra one may 
place some confidence on the theoretical estimation for the levels
$0^+_2$, $4^+_1$ and $3^-_1$.

\begin{table*}
\begin{ruledtabular}
\caption{\label{table.18C} Like table \ref{table.14C} for $^{18}$C.}
\begin{tabular}{ccccccc}
 Config             & $\nu$ & State     & $N_{\rm pair}$ & $E_i[MeV]$                  & $E[MeV]$  & $Ex[MeV]$ \\
\hline
 $(1)^2(2)^2(3)^2$  & 0     & $0^+$     & 3          & $E_1=-10.495$               & $-20.394$ & $0$  \\
                    &       &           &            & $E_{2,3}=(-4.950;\pm1.262)$ &           &      \\
 $(1)^2(2)^2(3)(3)$ & 2     & $2^+,4^+$ & 2          & $E_1=-10.531$               & $-17.525$ & $2.869$\\
                    &       &           &            & $E_2=-4.809$                &           &      \\
 $(1)^2(3)^2(2)(3)$ & 2     & $2^+,3^+$ & 2          & $E_1=-10.543$               & $-17.203$ & $3.191$\\
                    &       &           &            & $E_3=-3.710$                &           &      \\
 $(1)^2(3)^4$       & 0     & $0^+$     & 3          & $E_1=-10.549$               & $-17.020$ & $3.374$ \\
                    &       &           &            & $E_{2,3}=(-3.236;\pm0.474)$ &           &      \\
 $(1)^2(3)^2(3)(3)$ & 2     & $2^+,4^+$ & 2          & $E_1=-10.573$               & $-15.388$ & $5.006$ \\
                    &       &           &            & $E_3=-2.630$                &           &      \\
 $(2)^2(3)^2(1)(3)$ & 2     & $2^-,3^-$ & 2          & $E_{2,3}=(-4.177;\pm0.772)$ & $-14.393$ & $6.001$\\
 $(3)^4(1)(2)$      & 2     & $0^-,1^-$ & 2          & $E_{2,3}=(-3.576;\pm0.981)$ & $-13.955$ & $6.439$\\
 $(2)^2(3)^4$       & 0     & $0^+$     & 3          & $E_2=-4.405$                & $-11.659$ & $8.735$\\
                    &       &           &            & $E_{1,3}=(-3.627;\pm1.433)$ &           &     \\
 $(2)^2(3)^2(3)(3)$ & 2     & $2^+,4^+$ & 2          & $E_2=-4.008$                & $-9.347$  & $11.047$\\
                    &       &           &            & $E_3=-3.243$                &           &      \\
\end{tabular}
\end{ruledtabular}
\end{table*}

\begin{table*}
\caption{\label{table.20C} Like table \ref{table.14C} for $^{20}$C.}
\begin{ruledtabular}
\begin{tabular}{ccccccc}
 Config                  &$\nu$ &State     &$N_{\rm pair}$ &$E_i[MeV]$                 &$E[MeV]$  &$Ex[MeV]$ \\
\hline
 $(1)^2(2)^2(3)^4$       & 0    &$0^+$     &4          &$E_1=-10.379$              &$-22.194$ &$0$\\
                         &      &          &           &$E_2=-4.502$               &          &  \\
                         &      &          &           &$E_{3,4}=(-3.667;\pm1.546)$&          &  \\    
 $(1)^2(2)^2(3)^2(3)(3)$ & 2    &$2^+,4^+$ &3          &$E_1=-10.369$              &$-19.813$ &$2.381$\\
                         &      &          &           &$E_2=-3.993$               &          &  \\
                         &      &          &           &$E_3=-3.238$               &          &  \\
 $(1)^2(3)^4(2)(3)$      & 2    &$2^+,3^+$ &3          &$E_1=-10.406$              &$-19.047$ &$3.147$\\
                         &      &          &           &$E_{3,4}=(-2.846;\pm0.665)$&          &  \\    
 $(2)^2(3)^4(1)(3)$      & 2    &$2^-,3^-$ &3          &$E_2=-4.080$               &$-15.969$ &$6.225$\\
                         &      &          &           &$E_{3,4}=(-2.925;\pm0.896)$&          &  \\    
 $(3)^6(1)(2)$           & 2    &$0^-,1^-$ &3          &$E_4=-3.167$               &$-15.201$ &$6.993$\\
                         &      &          &           &$E_{2,3}=(-2.615;\pm1.113)$&          & 
\end{tabular}
\end{ruledtabular}
\end{table*}


\subsection{Results: Complex Energy Representation}
The first step in the determination of the complex representation is to
find the 
resonant partial waves. This is done by evaluating the outgoing  
solutions (Gamow states) of the Schrodinger 
equation \cite{1928Gamow,1968Berggren} of the mean field Hamiltonian
defined in 
Sec. \ref{sec.sp}. Then, the half-life of the Gamow state is compared with the characteristic time of the system $\tau_c=5.953 \times 10^{-23}$ sec (see 
Sec. \ref{sec.tau}). Table \ref{table.tau} compares the characteristic time 
with the half-life of the states $\varepsilon_{0d_{3/2}}$ and 
$\varepsilon_{0f_{7/2}}$. The half-life of the state $0d_{3/2}$ is around 
nine times bigger than the characteristic time. The $0f_{7/2}$ state seems to 
be a wide resonance, but the comparison with the characteristic time shows that 
its half-life is a bit bigger than $\tau_c$. 

\begin{table}[ht]
\caption{\label{table.tau}Comparison of the half-life versus the 
characteristic time (Sec. \ref{sec.tau}).}
\begin{ruledtabular}
\begin{tabular}{ccc}
 state      & $T_{1/2}$ [sec]         & $T_{1/2} / \tau_c$\\
\hline
 $0d_{3/2}$ & $5.485 \times 10^{-22}$ & $9.21$ \\
 $0f_{7/2}$ & $7.505 \times 10^{-23}$ & $1.26$
\end{tabular}
\end{ruledtabular}
\end{table}

\begin{figure}[ht]
\vspace{6mm}
\includegraphics[width=0.45\textwidth]{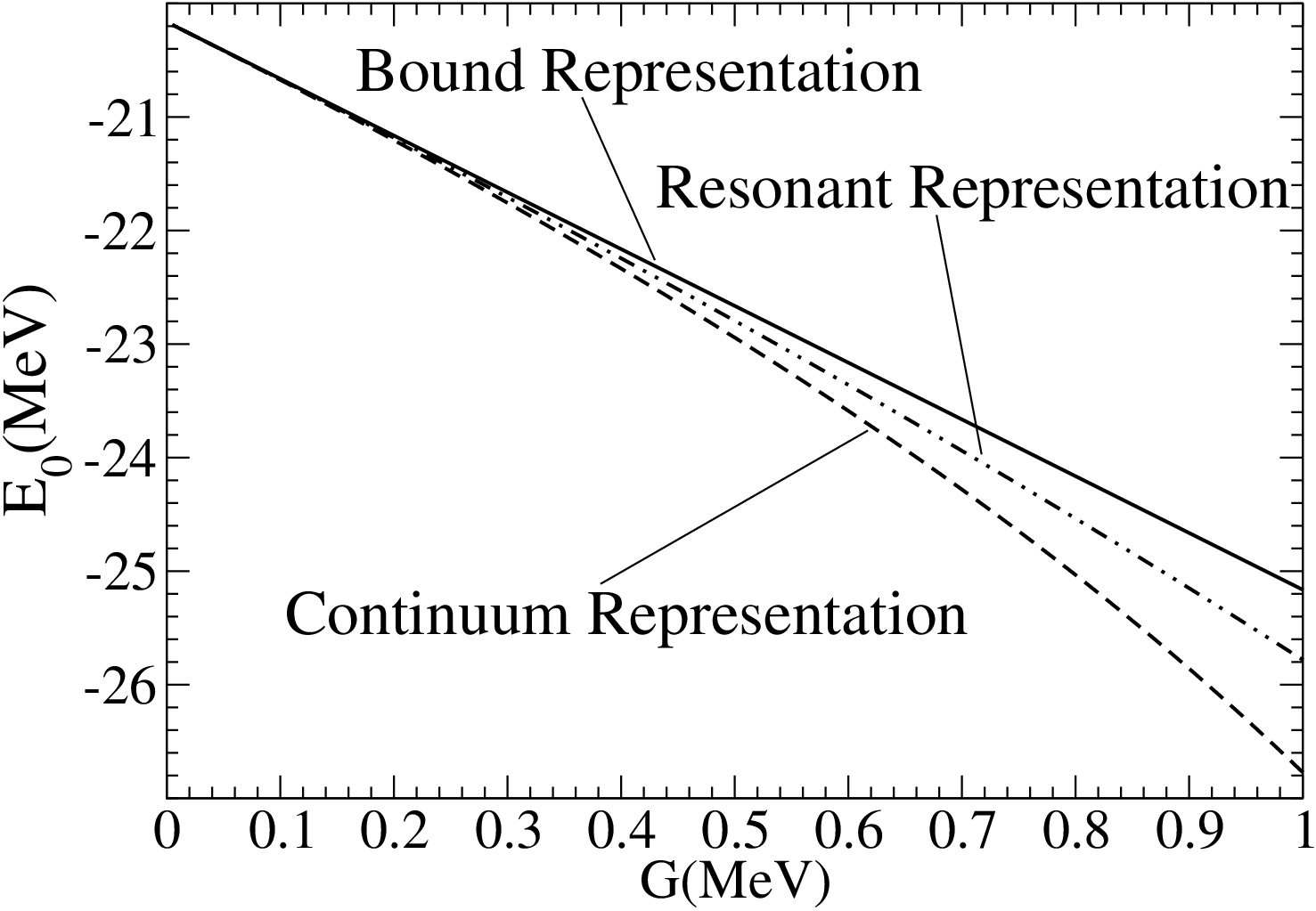}
\caption{\label{fig.22Ccxmpl} Ground state energy of $^{22}$C versus pairing 
strength $G$ for three different model spaces.}
\end{figure}

The effect of the resonant continuum was already investigated in ref. 
\cite{2003Hasegawa}. In order to investigate the effect of the non resonant 
continuum on the many-body correlations we compare in fig. \ref{fig.22Ccxmpl} 
the ground state energy of the nucleus $^{22}$C as a function of the pairing 
strength for three different model spaces: (i) Bound: $\{ 0p_{1/2}, 1s_{1/2}, 
0d_{5/2} \}$, (ii) Resonant: $\{ 0p_{1/2}, 1s_{1/2}, 0d_{5/2}, 0d_{3/2}, 
0f_{7/2} \}$, and (iii) Continuum (Secc. \ref{sec.real}). It is observed that 
the resonant and non resonant continuum states can be neglected as long as 
the pairing force is not very strong \cite{2003Hasegawa}. As the interaction 
increases the continuum starts to be important. The curve labeled as ''Continuum 
Representation" gives the ground state energy when the resonant and non resonant 
continuum is included in the representation through the CSPLD. The figure shows clearly the 
energy gain due to the inclusion of the continuum. The curve labeled as ``Resonant Representation" gives the energy when only the resonant states are 
included in the representation. For very big strength the non 
resonant continuum becomes as important as the resonant continuum.

\begin{figure}[ht]
\vspace{6mm}
\includegraphics[width=0.43\textwidth]{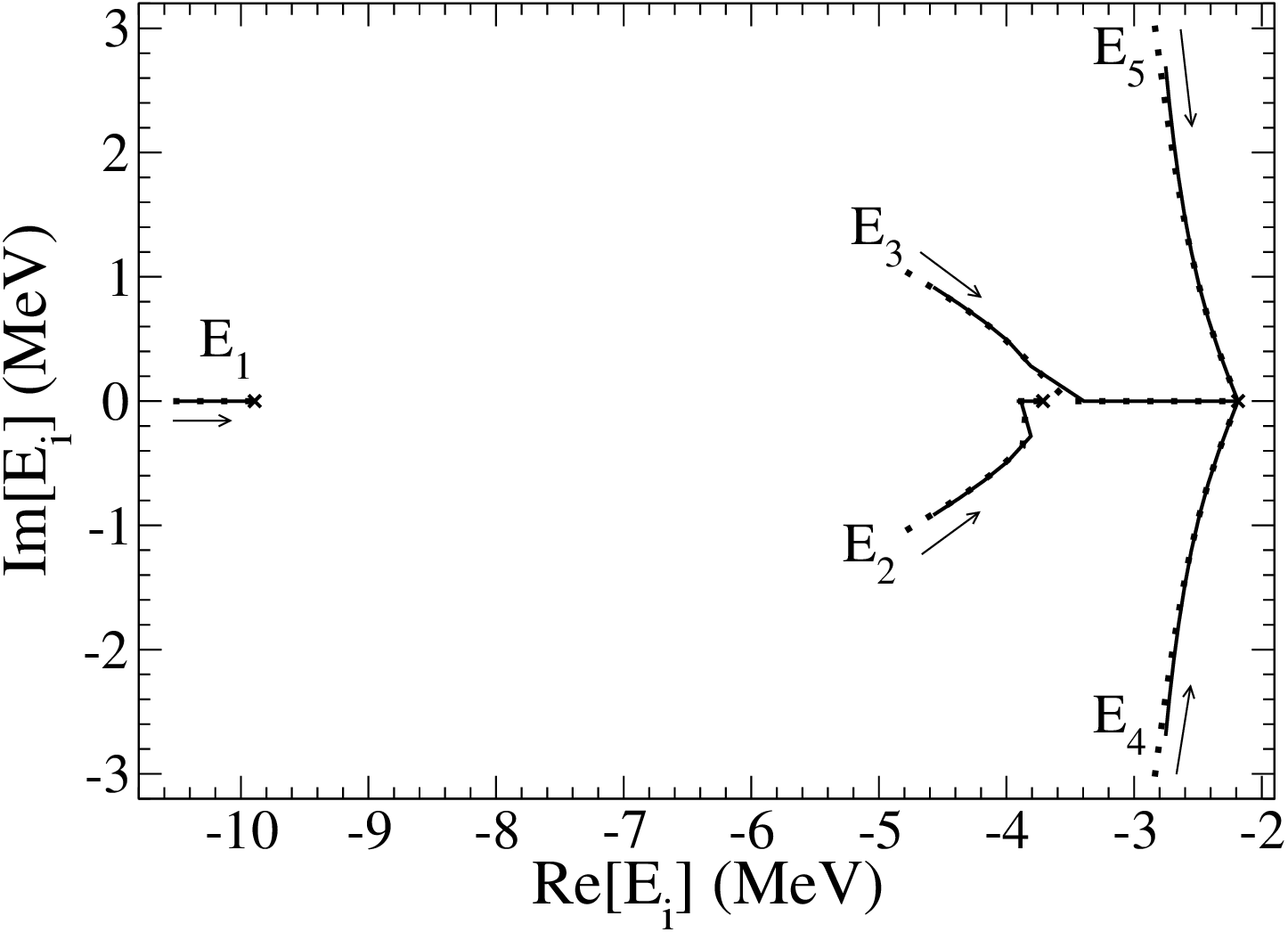}
\caption{\label{fig.cx-b} Pair energies in the ground state $^{22}$C versus 
pairing strength $G$ for $G=1.0$ MeV to $G=0.005$ MeV. The continuum line
corresponds to the bound representation while the dot line corresponds to the
resonant representation. 
The arrows point in the direction of decreasing $G$.}
\end{figure}

Let us compare the evolution of the pair energies $E_i$ in the bound and the 
resonant representation versus the pairing strength. Figure \ref{fig.cx-b} shows $E_i$ 
for $G$ from $G=1.0$ MeV to $G=0.005$ MeV in the nucleus $^{22}$C . The 
continuum (dot) line corresponds to bound (resonant) representation. The deeper 
pair energy $E_1$ is little affected by the model space (one can not 
distinguish between the two curves). The other pairs are more affected for 
big value of the strength. The difference diminishes 
as the interaction decreases. 
The same effect was observed in the ground state 
energy (fig. \ref{fig.22Ccxmpl}). The pairs $E_2$ and $E_3$ are complex 
conjugate partners for $G \gtrsim 0.51$ MeV and they move at the same pace as 
$G$ changes. When they become real $E_2$ approaches to the uncorrelated pair 
energy $2\varepsilon_2$ while $E_3$ moves faster to the uncorrelated pair 
energy $2\varepsilon_3$. The pairs $E_4$ and $E_5$ remain complex conjugate for 
all no zero values of the strength.

\begin{figure}[ht]
\vspace{6mm}
\includegraphics[width=0.45\textwidth]{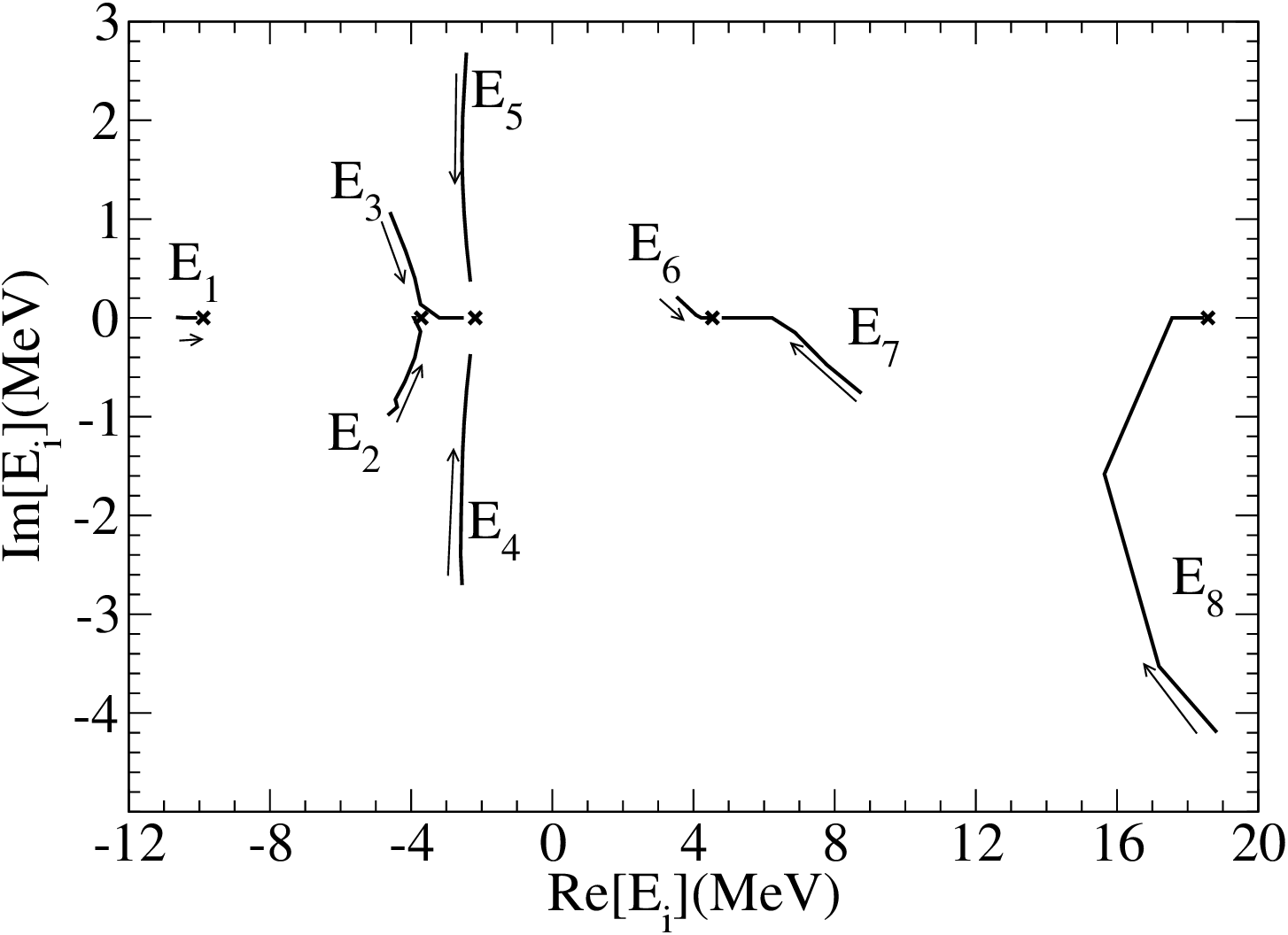}
\caption{\label{fig.28Ccxmpl} Evolution of the pair energies in the ground 
state of $^{28}$C as a function of the pairing strength $G$ from $G=2.2$ MeV 
to $G=0.2$ MeV. The arrows point in the direction of decreasing $G$.}
\end{figure}

As a last application we will calculate the evolution of pair energies in 
the continuum, i.e. pair energies with positive real component. To this aim 
let us study the nucleus $^{28}$C with eight pairs. Fig. \ref{fig.28Ccxmpl} 
shows the evolution of the pairs for strength from $G=2.2$ MeV to $G=0.2$ MeV. 
The bound (negative real component) pairs $E_1$ to $E_5$ follow a trajectory 
similar to that in $^{22}$C with the difference that the complex partners 
$E_2-E_3$ and $E_4-E_5$ are only approximately complex conjugate to each other. They become truly complex conjugate partners as the interaction approaches 
zero. On the other hand, the pairs in the continuum show a striking behavior. 
The typical movement to the right is not follow by all the positive energy 
pairs, i.e the continuum pairs may converge to its uncorrelated energy from 
right or left as $G$ decreases. Besides, the pairs seem to converge to the 
real part of the uncorrelated pair energy $lim_{G\rightarrow 0^+} E_i=2 
Re[\varepsilon_i]$ when $\varepsilon_i$ is a Gamow state.

\section{Conclusion} \label{sec.discussion}
The contribution of this paper to the exact solution of the pairing 
Hamiltonian is the inclusion of the resonant and non resonant continuum through 
the continuum single particle level density (CSPLD). The Gamow states, which 
appear in the complex energy representation, provide  
the main contribution from the continuum. It is
worthwhile to point out that in the representation
these states have exactly the same status as bound states. The difference
is that the states in the continuum are no affected by blocking effects.
The inclusion of the continuum has allowed us to 
study the unbound isotope $^{24}$C and beyond.
It was found that the continuum pairs (pair energies with positive real 
components) converge to the real part of the uncorrelated pair energy and they 
do not appear in complex conjugate partners. As a consequence the total energy 
may be complex. It was shown that from the exact solution of the pairing Hamiltonian 
the CSPLD can be used to investigate the effects of the resonant and non 
resonant continuum states upon the many-body pairing correlations.

\begin{acknowledgments}
This work has been partially supported by the National Council of Research 
PIP-77 (CONICET, Argentina).
\end{acknowledgments}

%

\end{document}